\font\caps=cmcsc10 at 12pt
\font\eightrm=cmr8 at 8pt
\documentclass{oxarticle}
\usepackage{tikz}

\usetikzlibrary{backgrounds,fit,decorations.pathreplacing,calc}
\usepackage{amsmath, amssymb}
\newcommand{\cconst}{cosmological constant}

\newcommand{\pas}{\slash \hspace{-.5em} \pa}

\newcommand{\cM}{{\cal M}}

\newcommand{\tr}{{transformation}}

\newcommand{\cG}{{\cal G}}

\newcommand{\cA}{{\cal A}}

\newcommand{\PB}{Master Equation}

\newcommand{\cL}{{\cal L}}
\newcommand{\bt}{\begin{tabular}{c}}
\newcommand{\et}{\end{tabular}}

\newcommand{\eb}{\ee\be } 
\newcommand{\ebp}{\rt.\ee\be\lt.} 
\newcommand{\bmat}{\lt ( \begin{array} }
\newcommand{\emat}{  \end{array} \rt )}

\newcommand{\oB}{{\ov B}}

\newcommand{\ovD}{{\ov D}}

\newcommand{\oE}{{\ov E}}

\newcommand{\ovY}{{\ov \Psi}}
\newcommand{\oc}{{\ov c}}

\newcommand{\ovG}{{\ov G}}

\newcommand{\oC}{{\ov C}}

\newcommand{\oF}{{\ov F}}
\newcommand{\A}{{\ov A}}
\newcommand{\B}{{\ov B}}

\newcommand{\Br}{\Bigr}
\newcommand{\Bl}{\Bigl}
\renewcommand{\a}{\alpha}	
\renewcommand{\b}{\beta}
\newcommand{\g}{\gamma}
\renewcommand{\d}{\delta}
\newcommand{\e}{\epsilon}
\newcommand{\ve}{\varepsilon}
\newcommand{\z}{\zeta}
\newcommand{\h}{\eta}
\newcommand{\q}{\theta}
\renewcommand{\k}{\kappa}
\newcommand{\lam}{\lambda}
\newcommand{\m}{\mu}
\newcommand{\n}{\nu}	
\newcommand{\x}{\xi}
	
\renewcommand{\r}{\rho}
\newcommand{\s}{\sigma}

\renewcommand{\t}{\tau}

\newcommand{\f}{\phi}
\renewcommand{\c}{\chi}
\newcommand{\y}{\psi}
\newcommand{\w}{\omega}

\newcommand{\Lam}{\Lambda}

\newcommand{\Y}{\Psi}

\newcommand{\la}{\label}
\newcommand{\ci}{\cite}

\newcommand{\ds}{\documentstyle}	
\newcommand{\fr}{\frac}

\newcommand{\pa}{\partial}
\newcommand{\ov}{\overline}
\newcommand{\be}{\begin{equation}}
\newcommand{\ee}{\end{equation}}
\newcommand{\ba}{\begin{array}} 
\newcommand{\ea}{\end{array}}
\newcommand{\bea}{\begin{eqnarray}}
\newcommand{\eea}{\end{eqnarray}}
\newcommand{\ra}{\rightarrow}
\newcommand{\Ra}{\Rightarrow}

\newcommand{\Lra}{\Leftrightarrow}
\newcommand{\lt}{\left}
\newcommand{\rt}{\right}

\newcommand{\ben}{\begin{enumerate}}
\newcommand{\een}{\end{enumerate}}
\newcommand{\bitem}{\begin{itemize}}

\newcounter{orange} 
\newcounter{apple} 
\addtocounter{apple}{1}

\newcounter{grape} 
\addtocounter{grape}{1}
\newcommand{\numberhere}{JB178}
\newcommand{\articlenumber}{Candidate}

\proofmodefalse
\usepackage{color} 

\begin{document}
%\pagestyle{empty}

%\vspace*{1in}

\renewcommand{\thefootnote}{\fnsymbol{footnote}}
%\footnotetext[1]{~here we have a footnote.}\renewcommand{\thefootnote}\arabicfootnote}}

%\vfill
%
%
 
 \begin{center}
{ \huge  A Candidate for a Supergravity Anomaly\\
[.5cm]} 
\vspace*{.1in}
{\caps John A. Dixon\footnote{cybersusy@gmail.com, \;john.dixon@ucalgary.ca}
\\ University of Calgary\\
Calgary, Alberta, Canada} \\[.5cm] 

\end{center}

\begin{center} Abstract  
\end{center}
%\large
\Large

\large

Using elementary BRS cohomology theory, this paper describes a supergravity anomaly analogous to, but very different from, the well known gauge and gravitational anomalies.   The form of the anomaly is simply the ghost charge one expression  
$\cA^{1}_{\rm Supergravity}= \int d^4 x\; \ve^{\lam\m\n\s} \lt \{ c_1 C_{ \a}
\pa_{\lam}  \Y_{\m}^{\a}   \pa_{\n}  V_{\s} + c_2
\pa_{\lam}  \Y_{\m}^{\a}     \pa_{\n} \Y_{\s \a} \; \w \rt \}
$.  Here $\Y_{\m\a}$ and $V_{\m}$ are the gravitino and a gauge boson, and $C_{\a}$ and $\w$ are their Faddeev Popov ghost fields. There is one boundary generator here, namely  the ghost charge zero expression   $\cA^{0}_{\rm Counterterm}= \int d^4 x\; \ve^{\lam\m\n\s} \lt \{\Y_{\lam  \a}
\pa_{\m}  \Y_{\n}^{\a}   V_{\s}\rt \}$, but there are two cycles in $\cA^{1}_{\rm Supergravity}$, so the anomaly is present if the coefficients $c_1,c_2$ are not in the ratio that can arise from the boundary generator. A model that is likely to generate this supergravity anomaly is described. The coefficient of this anomaly, in perturbation theory with unbroken supergravity, appears to be zero, because no relevant  diagrams are linearly divergent.  However,  when, and only when, there is spontaneously broken supergravity, there are counterterms needed in the action, and these counterterms are just right to contribute to linearly divergent diagams that can contribute to the anomaly.    We conjecture that the coefficient of the anomaly, for a general gauge group, is of the form $ \k^3 M_G \lt < D^a\rt > $, where $M_G$ is the mass of the gravitino, $\k$ is the Planck length, and $\lt < D^a\rt >$ is the VEV of the gauge auxiliary field.  This paper also discusses gauge parameters for the gravitino and the massive vector boson. It is possible that this anomaly might provide significant constraints to characterize viable theories  that are based on supergravity or superstring theory.

\Large
\refstepcounter{orange}
{\bf \theorange}.\;
{\bf Supergravity and Superstring Theory:}  These theories appeared likely to explain how we could generate a Grand Unified Theory, including supergravity, and then explain why the Standard Model looks so special and peculiar  \ci{{west},{freepro},{ferrarabook},{superspace},{WB},Buchbinder:1998qv}.  But the initial promise of these theories has yielded disappointment, because there does not seem to be any way to understand why one set of particles and groups is better than another.  At the same time supersymmetry  \ci{xerxes,Weinberg3,haberandallanach,buchmueller,ferrarasagnoti} has also been a disappointment,   because no plausible superpartners have been observed in the many   experiments that have looked for them.  
Anomalies in higher dimensional gravity theories generated much of the interest in superstring theory \ci{gsw1,gsw2}.  Gauge anomalies  play a very important role in the Standard Model \ci{pdgsm}. 

It could be argued that adding supergravity has not really helped in the search for a Grand Unified Theory.  Adding supergravity is very complicated. Incorporating it  might seem worthwhile, if only it explained something.  But, up to now, it has been generally believed that there are no further anomalies that are special to supergravity.  However, this paper claims that there are such anomalies.

 \refstepcounter{orange}
{\bf \theorange}.\;
{\bf The main result of this paper} is in   paragraph \ref{sugranomalyform}, and it is very simple to describe, as follows: There is at least one expression, with  ghost charge one, in the BRS cohomology of supergravity, which is not simply the supersymmetric version of any known gauge or gravitational anomaly,  and it is in equation (\ref{sugranomaly}) below. It is also true that the methods needed for the calculation of the coefficient of this anomaly are very different from those for any other known anomaly, as will be shown below, particularly in paragraphs \ref{diagramforanomalyparagraph} and \ref{newcounterterms}.

 \refstepcounter{orange}
{\bf \theorange}.\;
{\bf General Remarks about the BRS Cohomology of Supergravity:}  
 The questions regarding supersymmetry and BRS symmetry \ci{taylor, poissonbrak,{Becchi:1975nq}, {zinnbook},{Zinnarticle},{becchiarticles1},
becchiarticles2} are popular topics.  The subject, in general, is immense and confusing.    A particularly  important issue is that there are  various claims that the BRS cohomology of supergravity  and related theories have been examined, and that no ghost charge one objects, special to supersymmetry, exist. It seems to have been generally accepted that supergravity  does not have any anomaly that is not simply the supersymmetric version  of a well known non-supersymmetric anomaly.   This paper shows that this viewpoint is not correct.  At the end of this paper, there are some tables which summarize some of the notation and results in this paper.

\refstepcounter{orange}
{\bf \theorange}.\;
{\bf BRS and ZJ and the SU(3) and U(1) Anomalies:}  
The power of the Becchi Rouet Stora (BRS) and Zinn-Justin (ZJ) Master Equation technique is that it summarizes all the symmetries of an action, and its transformations, in a compact and powerful way.  Originally BRS used the mathematical fact  that if one takes
\be
\d=  \w^a  T^a- \fr{1}{2}f^{abc} \w^b \w^c \fr{\pa}{\pa \w^a}
\ee
where $T^a$ are the antihermitian generators of the SU(3) group on some  field, satisfying
\be
\lt [ T^a, T^b \rt ] =  f^{abc} T^c 
\ee
then this generates cohomology because $\d^2 = 0 $.  They noticed that the anomaly of the SU(3) Yang Mills theory in four dimensions arises from  the totally antisymmetric tensor with five ghosts (here $d^{abc}$ is the symmetric tensor with three indices in the real adjoint representation of SU(3)):
\be
d^{abc} f^{bde} f^{cpq} \w^a \w^d \w^e \w^p\w^q 
\eb
\ra \cA^{1}_{\rm SU(3)\; Anomaly}= \int d^4 xd^{abc}  \lt \{ \ve^{\m\n\lam\s}
\pa_{\m} V_{\n}^{a} \pa_{\lam} V_{\s}^{b} \w^c+ \cdots\rt \}
\la{SU3anomaly}
\ee 
The $+ \cdots $ in the above are higher order terms in the coupling constant and we can ignore them when computing the anomaly. The notation $ \cA^{1}_{\rm SU(3)\; Anomaly}$ is to remind us that this expression looks like a piece of the action, except that it has ghost charge one.

\refstepcounter{orange}
{\bf \theorange}.\;
{\bf   Similarly the usual U(1) anomaly arises from the expression:}
\be
\cA^{1}_{\rm U(1)\;Anomaly}= \int d^4 x \lt \{ \ve^{\m\n\lam\s}
\pa_{\m} V_{\n} \pa_{\lam} V_{\s} \w \rt \}
\la{U1anomaly}
\ee 
  The Grassmann odd ghost fields  $\w^a, \w$ are the Faddeev Popov (FP) Ghosts. Early on, ZJ introduced a way of incorporating the nilpotent $\d$ of BRS  into the path integral technique. Later on, Batalin and Vilkovisky (BV) named  these  results the `Master Equation' \ci{{BV},{Weinberg2}}.

 \refstepcounter{orange}
{\bf \theorange}.\;
{\bf Gauge Transformations:}  
These anomalies arise from the following nilpotent gauge transformations:
\be
\d V^a_{\m} =  \pa_{\m} \w^a + \cdots
\ee
\be
\d V_{\m} =  \pa_{\m} \w + \cdots
\ee
These simple terms are nilpotent because they are Grassmann odd.  When the higher order terms are included, it gets more complicated.

 \refstepcounter{orange}
{\bf \theorange}.\;
{\bf Why is the expression an anomaly?}  
Here is the reason. The expressions like (\ref{U1anomaly}) 
 can be obtained,  in perturbation theory,
by taking the BRS variation of some Feynman diagram one-particle-irreducible vertex, as follows, for example (see  (\ref{anomdiagram}) below for an example of the calculation):
\be
\cA^{1}_{\rm U(1)\;Anomaly}= \d {\cG}^{0}_{\rm Feynman}
\ee
This vertex $ {\cG}^{0}_{\rm Feynman}$ is a complicated non-local expression and the calculation is an interesting and  subtle process, which necessarily uses the linear divergence of the amplitude ${\cG}^{0}_{\rm Feynman}$.  See \ci{taylor} for a simple discussion of this process. 
 Clearly one can try to remove the above expression (\ref{U1anomaly}) as follows:
\be
\cA^{1}_{U(1)}= \d \cA^{0}_{\rm U(1)\;Counterterm}
\la{anomisaboundary}
\ee
where one chooses a local counterterm in the action, of the form:
\be
\cA^{0}_{\rm U(1)\;Counterterm}= \fr{1}{2} \int d^4 x \lt \{ \ve^{\m\n\lam\s}
V_{\m} V_{\n} \pa_{\lam} V_{\s}  \rt \}
\la{U1counterterm}
\ee 
The problem is that this particular expression $\cA^{0}_{\rm U(1)\;Counterterm}=0$.  
We are antisymmetrizing a symmetric expression $V_{\m} V_{\n}$.
In fact there is no appropriate local expression that can satisfy equation (\ref{anomisaboundary}). Exactly the same problem arises for expression (\ref{SU3anomaly}). Our example of a supergravity anomaly (sugranomaly) below in
(\ref{sugranomaly}) 
 has exactly the same features, but it is trickier because it involves the gravitino.

 \refstepcounter{orange}
{\bf \theorange}.\;
{\bf BRS realized that anomalies can be identified as objects in the local BRS ghost charge one cohomology:}   The reason for this is simple.  We observe that for example, it is true that for the relevant nilpotent BRS operators $\d$, the objects in equations  
(\ref{U1anomaly}) and 
(\ref{SU3anomaly}) satisfy the equations
\be
\d 
 \cA^{1}_{\rm SU(3)\; Anomaly}=
\d 
 \cA^{1}_{\rm U(1)\; Anomaly}=0
\ee
but there do not exist any local polynomial objects such that 
\be
 \cA^{1}_{\rm SU(3)\; Anomaly}= \d \cA^{0}_{\rm SU(3)\; Counterterm}
\ee
\be 
 \cA^{1}_{\rm U(1)\; Anomaly}=\d\cA^{0}_{\rm U(1)\; Counterterm}
\ee
In practice,   the expressions $\cA^{1}_{\rm SU(3)\; Anomaly}$ do arise in perturbation theory, by taking the variation of some Green's function.  If the anomaly is in the local BRS cohomology space, it cannot be removed by any terms added to the Lagrangian, since the latter is local and the 1PI vertices are not local.  This is true at the lowest level, and it continues to be true at all levels in the relevant coupling constants.

 \refstepcounter{orange}
{\bf \theorange}.\;
{\bf Supergravity Anomaly:}  
\la{sugranomalyform}
This paper will argue that  supergravity has an anomaly of its own, of the form
\be
\cA^{1}_{\rm Supergravity}= \int d^4 x\; \ve^{\lam\m\n\s} \lt \{ c_1 C_{ \a}
\pa_{\lam}  \Y_{\m}^{\a}   \pa_{\n}  V_{\s} + c_2
\pa_{\lam}  \Y_{\m}^{\a}     \pa_{\n} \Y_{\s \a} \; \w + \cdots\rt \}
\la{sugranomaly}
\ee 
where the numerical coefficients $c_i,i=1,2$ are discussed below. 
In the above formula, $\Y_{\m \a}$ is the Grassmann odd spin $\fr{3}{2}$  gravitino of supergravity (in the complex two-component formulation), and $V_{\m}$ is a (real) vector boson in a gauge multiplet coupled to chiral multiplets and to supergravity.  The field $C_{\a}$ is the Grassmann even FP ghost for the supersymmetry and these two fields have the nilpotent transformations:
\be
\d \Y_{\m \a} = M_P \pa_{\m} C_{\a}+ \cdots
, \;
\d V_{\m} =  \pa_{\m} \w + \cdots
\la{theothersimpledelta}\ee
where $M_P$ is the Planck mass, and $\w$ is again the Grassmann odd scalar FP ghost, as in the above examples.  For the purposes of calculating the coefficient of the anomaly the higher order terms $+\cdots $ in equation (\ref{sugranomaly}) and equation (\ref{theothersimpledelta}) can be ignored, and that is very important for a complicated calculation like this.  

 \refstepcounter{orange}
{\bf \theorange}.\;
\la{sugracounter1}
{\bf  The counterterm situation for supergravity is more complicated than for the gauge theories:}
The relevant possible counterterm like (\ref{U1counterterm}) 
 would be\be
\cA^{0}_{\rm Supergravity\;Counterterm\;1}= \int d^4 x \lt \{ \ve^{\m\n\lam\s}
 \Y_{\m}^{\a}  \Y_{\n \a} \pa_{\lam} V_{\s}  \rt \}=0
\la{sugracounternumber1}
\ee 
and again this is zero, which is just like the situation for (\ref{U1counterterm}), except that here there is a triple antisymmetrization which makes this zero, whereas in (\ref{U1counterterm}) it is a single antisymmetrization.  However, 
 there is another possible counterterm in this supergravity case.
This takes the form
\be
\cA^{0}_{\rm Supergravity\;Counterterm\;2}= \int d^4 x \lt \{ \ve^{\m\n\lam\s}
\pa_{\lam}  \Y_{\m}^{\a}  \Y_{\n \a}  V_{\s}  \rt \}
\neq 0
\la{secondsugracounter}
\ee 
An effort to show that this is zero results in a quadruple antisymmetrization (including integration by parts), and so it does not succeed in showing this is zero.
This expression  generates objects that look exactly like  (\ref{sugranomaly}):

\large
\be
\d \cA^{0}_{\rm Supergravity\;Counterterm\;2}
= \int d^4 x\; \ve^{\lam\m\n\s} \lt \{ c'_1C_{ \a}
\pa_{\lam}  \Y_{\m}^{\a}   \pa_{\n}  V_{\s} + c'_2
\pa_{\lam}  \Y_{\m}^{\a}     \pa_{\n} \Y_{\s \a} \; \w \rt \}
\la{sugranomalyboundary}
\ee
\Large
where we have integrated by parts.  The values are $M_P c'_1= -c'_2= 1$.
Note that this combination of these two factors  occurs with a definite and unchangeable linear combination with definite numbers $c'_1, c'_2$ in the above expression (\ref{sugranomalyboundary}). An effort to show that the second term on the right hand side  in (\ref{sugranomalyboundary}),
 is zero, results in a double  antisymmetrization, and so the term is non-zero. 
The term in (\ref{secondsugracounter}) can generate one particular linear combination of these numbers, but not two. Both of the expressions in $\cA^{1}_{\rm Supergravity}$ in  (\ref{sugranomaly}) separately yield zero when acted upon by $\d$.   So there is no way, in general,  to obtain expression (\ref{sugranomaly}) from the BRS variation of a local object with ghost charge zero in the theory.  If it arises in perturbation theory with  coefficients that do not have the same ratio as in the above expression (\ref{sugranomalyboundary}), it will spoil the SUSY type gauge symmetry of supergravity in an incurable way.   In quantum field theory, we expect that this will happen in general--the rule there is that what can happen, usually does happen, because the perturbation expansion tends to explore all the corners. 

 \refstepcounter{orange}
{\bf \theorange}.\;
{\bf Equations of motion of the gravitino and the vector boson:}
We will not  worry here about the issues that relate to the higher levels in the coupling.  However, it is crucial to observe that a genuine anomaly must not vanish using the equations of motion.  Sometimes this is called invariance under deformations of the transformation rules \ci{Imbimbo}.  This requirement is equivalent to  ensuring that the anomaly is  in the cohomology space (a cocyle, but not a coboundary) of the entire $\d$ derived in paragraph \ref{mastereq} here. The equation of motion for the gravitino is discussed  below in paragraph  \ref{gravitinopropagatorterm}. The expression (\ref{sugranomaly}) does not vanish using any of the forms of the equations of motion of either the gravitino or the vector boson.

 \refstepcounter{orange}
{\bf \theorange}.\;
{\bf The supergravity anomaly requires the spontaneous breaking of supersymmetry in order  to appear in the theory:}  Later in this paper, starting with paragraph \ref{starttowritedowntheaction}, 
 we will write down the detailed components of a theory which we believe will generate this anomaly.  Based on this theory, we present a diagram for the anomaly in (\ref{anomdiagram}).    The surprising thing here is that this anomaly of supergravity does not make its appearance in perturbation theory in the simple way familiar from the two anomalies of gauge theory above.  In fact, a preliminary examination of  the diagrams indicates that there are no linearly divegent diagrams for the relevant amplitude, which excludes any possibility of generating it with a non-zero coefficient. 
I have tried to find a linearly divergent diagram for the relevant amplitude in the theory here, before introducing the terms described in paragraph   \ref{newcounterterms},  and it does not seem possible, but I have no proof that it does not happen.

 \refstepcounter{orange}
{\bf \theorange}.\;
{\bf Conjecture for the  supergravity anomaly:} My  conjecture is, that for this anomaly to appear,  SUSY and gauge symmetry must both be spontaneously broken.  This means that the gravitino mass, and the vector boson mass, must both be present, and must  arise from the spontaneous breaking of local supersymmetry and local gauge symmetry, using  the well known Higgs mechanisms \ci{higgs}. 
The reasoning behind this conjecture follows from the diagram (\ref{offdiagonalprop}) 
 in paragraph \ref{newcounterterms} below.  At the present time, this claim remains a conjecture.

\refstepcounter{orange}
{\bf \theorange}.\;
{\bf  A diagram for the sugranomaly:}
\la{diagramforanomalyparagraph}
Here is an example of a  diagram that appears likely to generate this anomaly in supergravity. It is a linearly divergent triangle diagram, which is a necessary feature for a diagram to generate an anomaly, by analogy with the known anomalies discussed above.
However, there is  a very serious problem in this diagram.  It needs a propagator
from ${\ov \lam}^{\dot \g}$ to $\c_1^{\d_1}$ in the top left of the triangle (we are using a circle, actually), but no such propagator exists in the theory!  This problem appears, at first sight, to be an insuperable difficulty for generating the anomaly. As we shall see below, this problem is actually a hint about how the anomaly can be generated.

\be
\begin{picture}(400,280)
\put(100,0){Diagram 1 for $\cG\lt[ \Y,\Y,V \rt ]$}
%
%\put(0,250){ From vertices $$}
%
\put(0,250){ $ \cG_{1}\lt[ \Y,\Y,V \rt ]=\k^2  \int d^4 k\; d^4 p \;d^4 q  \;\d^4(k + p + q)\Y^{\m \a}(p)  \Y^{\n\b}(q)  V^{ \r}(k) $ }

\put(0,215){ $
 {\displaystyle
 \fr{1}{(2 \pi)^4} \int   d^4 l  \lt [\s_{\n_1 \a\dot \g} (l+q)_{\m} -\s_{\m \a\dot \g} (l+q)_{\n_1}    \rt]
\lt ( \fr{  (l-k)\cdot\ov \s }{  (l-k)^2 +M_{\rm 1,mix}^2 }  \rt )^{ \dot \g \d} 
}
$ }

\put(0,170){ $
 {\displaystyle 
i g   \s_{\r \d \dot \ve} \fr{  M_{\rm 1,mix} \ve^{\dot \ve \dot \z} }{  l^2  + M_{\rm 1,mix}^2}  
  \lt [\s_{\n_2 \b\dot \z} (l+q)_{\n} -\s_{\n \b\dot \z} (l+q)_{\n_2}    \rt]
\lt ( \fr{ \h^{\n_1 \n_2} + \cdots}{ (l+q)^2+ M_V^2}  \rt )  } 
$ }
%
%
%\put(0,150){ $ integral $  }
%
% four point at left:
%\put(180,75){\line(-1,-1){50}}
%\put(180,75){\line(-1,1){50}}
% top oblique at right:
\put(210,92){\line(1,2){15}}
% bottom oblique at right:
\put(210,58){\line(1,-2){20}}
% horizontal at right:
%\put(220,75){\line(1,0){70}}
% four point at right:
%\put(220,75){\line(1,1){50}}
%\put(220,75){\line(1,-1){50}}
% horizontal at left:
\put(180,75){\line(-1,0){70}}
\put(200,75){\circle{40}}
%
% outer fields
%a1
%\put(95,20){${\ov a}_1(p)$}
%a2
%\put(95,120){${\ov a}_2(p)$}
%a3
\put(230,120){$V^{\r }(k)$}
%a4
%\put(280,120){${\ov a}_4(p)$}
%a5
%\put(285,85){${\ov a}_5(p)$ }
%a6
%\put(280,20){${\ov a}_6(q)$}
%a7
\put(240,20){$\Y^{\n\b}(q)$}
% inner Fields:
%w
%\put(50,125){{$\Y(\fr{p}{2}+r)$}}
\put(50,75){{$\Y^{\m \a}(p)$}}
%\put(50,25){{$\Y(\fr{p}{2}-r)$}}
%b1
%\put(175,100){$\c$}
%b2
\put(155,85){ $\ov\lam^{\dot \g}$}
%b4
\put(192,103){${\c}_{1}^{\d}$}
%b5
\put(220,88){${\ov \c}_1^{\dot \ve}$}
%b7
%\put(185,42){$\c$}
%b8
%=
\put(160,51){$V^{\n_1}$ }
\put(185,38){$V^{\n_2} $ }
% masses
%\put(186,88){$\ast$}
%\put(184,81){${m }$}
%m2
%\put(207,89){$\ast$}
%\put(199,84){${m}$}
%m3
\put(216,70){$\ast$}
\put(225,72){$ M_{\rm 1,mix}$}
%m4
%\put(207,55){$\ast$}
%\put(200,64){${m}$}
%m5
%\put(186,56){$\ast$}
\put(220,53){${\ov \lam}^{\dot \z}$}
%m6
%\put(177,73){$\ast$}
%\put(185,73){${m}$}
%
% momenta
%{l}+{k}
\put(120,105){$\nearrow$}
\put(130,130){$l-k$}
%l
\put(265,75){$\downarrow$}
\put(280,75){$l$}
%l+q
\put(125,35){$\nwarrow$}
\put(100,25){$l+q$}
%\put(340,90){$k+p+q=0$}
%
\end{picture} 
\la{anomdiagram}
\ee
\Large

The vertices for the above diagram can be found in the chiral action in paragraph  \ref{chiralA1action},  and in the gauge action in paragaph \ref{U1action}.
However the new mixed kinetic term $ c_{\rm new}
\lam^{\a} \s^{\m}_{\a \dot \b}\pa^{\m} {\ov \c}_1^{\dot \b}$ arises from the new counterterm  (\ref{divergentmixedkinetic}) in the action, discussed in paragraph \ref{newcounterterms}. The massive $VV$ propagator  at the bottom  comes from the terms in paragraph \ref{vectorbosonpropagator}. Its mass could be dropped here, since only the linear divergence contributes to the anomaly. 

Note that there is no problem in getting the propagator $M_{\rm 1,mix} \ov \lam^{\dot \a} 
{\ov \c}_{1 \dot \a}  $ here, since it arises immediately from the term  $ i g   \ov \lam^{\dot \a} 
{\ov \c}_{1 \dot \a} \lt < A_{1}\rt >$ 
which is in the last line  (\ref{lastlineofchiralaction}) of the chiral action below. It  brings in a power of mass from the VEV. It is, of course, unusual to find a power of mass in an anomaly, but this diagram clearly has a power of $ g M_O v = M_{\rm 1,mix}$ from one propagator.

\refstepcounter{orange}
{\bf \theorange}.\;
{\bf  Calculation of the sugranomaly:}  We now recall that the sugranomaly is present in the theory if the following equation holds true in perturbation theory:
\be
\d  \cG\lt[ \Y,\Y,V \rt ]=  \int d^4 x\; \ve^{\m\n\lam\s} \lt \{ 
c_1 \pa_{\lam}  \Y_{\m}^{\a} C_{ \a}  \pa_{\n}  V_{\s} + c_2
\pa_{\lam}  \Y_{\m}^{\a}     \pa_{\s} \Y_{\n \a} \; \w \rt \}
\ee
where the vertex is the sum of all relevant diagrams, as usual:
\be
 \cG\lt[ \Y,\Y,V \rt ]=\sum_n \cG_{n}\lt[ \Y,\Y,V \rt ]
\ee

Here $c_1$ and $c_2$ are  coefficients, calculated from perturbation theory.  In order for this to be an anomaly, it is necessary that these coefficients not be in a ratio such that they can arise from  equation (\ref{sugranomalyboundary}). It is quite remarkable that this requires us to have a propagator resulting from a counterterm that arises from diagrams like 
(\ref{offdiagonalprop}) below.  Note that by the usual interpretation of quantum field theory, this term appears with a new arbitrary coefficient $ c_{\rm new}$, as further explained below in paragraph \ref{generator}.

\refstepcounter{orange}
{\bf \theorange}.\;
\la{newcounterterms}
{\bf  The  Counterterms for the theory with spontaneously broken SUSY:}  As we mentioned above, the problem with the diagram in  paragraph \ref{diagramforanomalyparagraph} is that there are no propagators that take ${\lam}^{\g_1}$ to ${\ov \c}_1^{\dot \d_1}$ in the theory.  But consider the following diagram:

\be
%template-improved March 1, 2000:
\begin{picture}(400,280)
\put(100,0){Diagram 1 for $\cG\lt[ \lam, \ov \c \rt ]$}
%
%\put(0,250){ From vertices $$}
%
\put(0,250){ $ \cG_1\lt[ \lam, \ov \c \rt ]= \int   d^4 p  \lam^{ \a}(p)  \ov \c_1^{\dot \b}(-p)    \;  \fr{1}{(2 \pi)^4} \int   d^4 l$ }

\put(0,215){ $
 {\displaystyle
\k^2 ( i g M_O v  ) ( M_G)  \lt ( \fr{  \ve^{\dot \g \dot \e} \h^{\n\m}+ \cdots }{  l^2  + M_G^2}  \rt )   
        }
$ }

\put(0,170){ $
 {\displaystyle 
  \lt [\s_{\n_1 \a\dot \g} (l+p)_{\n} -\s_{\n \a\dot \g} (l+p)_{\n_1}    \rt]
\lt ( \fr{ \h^{\n_1 \n_2} + \cdots}{ (l+p)^2+ M_V^2}  \rt )   ({\ov \s}_{\m  \n_2})_{\dot \ve \dot \b}} 
$ }
%
%
%\put(0,150){ $ integral $  }
%
% four point at left:
%\put(180,75){\line(-1,-1){50}}
%\put(180,75){\line(-1,1){50}}
% top oblique at right:
%\put(210,92){\line(1,2){15}}
% bottom oblique at right:
%\put(210,58){\line(1,-2){20}}
% horizontal at right:
\put(220,75){\line(1,0){70}}
% four point at right:
%\put(220,75){\line(1,1){50}}
%\put(220,75){\line(1,-1){50}}
% horizontal at left:
\put(180,75){\line(-1,0){70}}
\put(200,75){\circle{40}}
%
% outer fields
%a1
%\put(95,20){${\ov a}_1(p)$}
%a2
%\put(95,120){${\ov a}_2(p)$}
%a3
%\put(230,120){$V^{A\r }(k)$}
%a4
%\put(280,120){${\ov a}_4(p)$}
%a5
%\put(285,85){${\ov a}_5(p)$ }
%a6
%\put(280,20){${\ov a}_6(q)$}
%a7
%\put(240,20){$\Y^{\n\b}(q)$}
% inner Fields:
%w
%\put(50,125){{$\Y(\fr{p}{2}+r)$}}
\put(70,75){{$\lam^{\a}(p)$}}
\put(300,75){{${\ov \c}_1^{\dot \b}(-p)$}}
%\put(50,25){{$\Y(\fr{p}{2}-r)$}}
%b1
%\put(175,100){$\c$}
%b2
\put(160,85){ ${\ov \Y}^{\n \dot \g}$}
%\put(190,95){ ${\ov \Y}^{\m \d}$}
%b4
\put(192,103){$M_G$}
\put(195,92){$\ast$}
%b5
\put(220,88){${\ov \Y}^{\m  \dot \ve}$}
%b7
%\put(185,42){$\c$}
%b8
%=
\put(160,51){$V^{\n_1}$ }
%\put(185,38){$A^G$ }
% masses
%\put(186,88){$\ast$}
%\put(184,81){${m }$}
%m2
%\put(207,89){$\ast$}
%\put(199,84){${m}$}
%m3
\put(216,71){$\ast$}
\put(181,72){\small${\k i g M_O v}$}
%m4
%\put(207,55){$\ast$}
%\put(200,64){${m}$}
%m5
%\put(186,56){$\ast$}
\put(220,53){${V}^{\n_2}$}
%m6
%\put(177,73){$\ast$}
%\put(185,73){${m}$}
%
% momenta
%{l}+{k}
%\put(120,105){$\nearrow$}
%\put(130,130){$l-k$}
%l
%\put(240,75){$\downarrow$}
%\put(255,75){$l$}
%l+q
%\put(125,35){$\nwarrow$}
%\put(100,25){$l+q$}
%\put(340,90){$k+p+q=0$}
\put(195,35){$\rightarrow$}
\put(195,25){$l+p$}
\put(195,115){$\leftarrow$}
\put(195,125){$l$}
\end{picture} 
\la{offdiagonalprop}
\ee
\Large

The vertex at the left comes from  the action $\cA_{\rm Gauge}$ in   paragraph \ref{U1action}.   Similarly, the  vertex at the right comes from     the action $\cA_{A_1}$ in paragraph  \ref{chiralA1action} below after the VEV $<A_1>= M_O v$ described in paragraph \ref{vevvaluesparagraph} appears. There is a similar term and diagram from $\cA_{A_2}$. 
The propagator  at the top  comes from the action in paragraph \ref{gravitinopropagatorterm}.
There is still some work needed to convert that to the massive case. 
The propagator  at the bottom  comes from the terms in paragraph \ref{vectorbosonpropagator}.  This diagram clearly vanishes if the Gravitino mass $M_G$
vanishes.

\refstepcounter{orange}
{\bf \theorange}.\;
\la{countertermsformssgravitino}
{\bf  The Diagram (\ref{offdiagonalprop}) can exist only if the gravitino is massive:} 
Otherwise the propagator ${\ov \Y}^{\n_1 \dot \g_1} {\ov \Y}^{\n_2 \dot \g_2}$ does not exist.  This diagram is linearly divergent, and we can expect that it will need a counterterm in our action.  But this counterterm will be needed  only if the SUSY symmetry is spontaneously broken.  We expect that the calculations in our theory will satisfy the Master Equation (see below in equation (\ref{masterequation})), which reflects the complicated symmetry of this action.  What that means is that if there are counterterms needed in the theory, like the following
\be \cA_{\rm A,1\;new\;counterterm\;needed}=
c_{\rm new} \int d^4 x \; \lt \{
\lam^{\a} \s^{\m}_{\a \dot \b} \pa_{\m} {\ov \c}_1^{\dot \b}\rt \}
\la{divergentmixedkinetic}
\ee
where $c_{\rm new}$ is the divergent coefficient calculated from diagrams like  (\ref{offdiagonalprop}),  then there will also be other counterterms needed with the same coefficient. These can all be generated together as a boundary, as will be discussed below in paragraph \ref{generator}, after we have discussed more of the details of the Master Equation and actions here.

Note that $c_{\rm new}$ can indeed be dimensionless, because  the counterterm with a divergence in the diagram above in 
(\ref{offdiagonalprop}) for  $\cG_1\lt[ \lam, \ov \c \rt ]$ is proportional to $\int d^4 x \k^2 M_O M_G \lam^{\a} \s^{\m}_{\a \dot \b} \pa_{\m} {\ov \c}_1^{\dot \b}$ times a dimensionless (but divergent) constant, so that the whole term is of dimension zero.  On the other hand, we also note that the counterterm with a divergence in the diagram for  $ \cG_{1}\lt[ \Y,\Y,V \rt ]$ has dimension $\int d^4 x \k^2  M_{\rm 1,mix} \Y \pa \Y  V$, 
which is also of dimension zero.  This means that we can expect to calculate the sugranomaly coefficients in the form $c_1 = \k M_{\rm 1,mix} {\rm N}_1,c_2 = \k^2 M_{\rm 1,mix} {\rm N}_2$, where ${\rm N}_i$ are finite, but mass-dependent, numbers.  See paragraph \ref{conjectureforcoefficient} for more comments on these issues.

\refstepcounter{orange}
{\bf \theorange}.\;
\la{boundarycounterterms}
{\bf The   counterterms like (\ref{divergentmixedkinetic}), calculated from diagrams like (\ref{offdiagonalprop}), which fall into groups as discussed in paragraph \ref{generator},  must be added to the action when the gravitino is massive:} They then allow the diagrams like (\ref{anomdiagram}) to appear in the theory, and so they allow the sugranomaly to arise. That is why we said in paragaph 
\ref{diagramforanomalyparagraph} that the strange propagators needed for that   diagram are a hint as to how it arises.

\refstepcounter{orange}
{\bf \theorange}.\;
\la{diagonalizingtheterms}
{\bf In a more detailed analysis of this situation, one would want to diagonalize the various terms,} so that one is dealing with propagators that are diagonal in the masses and in the kinetic terms.  Here there is all sorts of mixing going on, and it takes a lot of effort to sort it out.  Our claim is that we can still find a lot of information from the action without doing all that work.  But that work now needs to be done to determine how these anomalies restrict the set of viable supergravity theories, particularly in the context of unification \ci{pran,ross,GUT}.

\refstepcounter{orange}
{\bf \theorange}.\;
\la{arealsugraanomaly}
{\bf Is this a real supergravity anomaly?}  The reader might wonder whether this is just an example of one of the usual anomalies, but made supersymmetric.  It seems clear that this is actually quite different from the known anomalies, although it certainly has a family resemblance, as noted in this paper.  We can expect that eliminating it from theories will give rise to interesting new restrictions on viable theories.  However, getting its general form, performing the Feynman calculations accurately, and figuring out what those restrictions are, are complicated and demanding tasks, given the complicated nature of the action and transformations.

 \refstepcounter{orange}
{\bf \theorange}.\;
\la{conjectureforcoefficient}
{\bf Conjecture for the Coefficient of the Anomaly in General:}  

For the general case where we have an arbitrary gauge group and arbitrary representations for matter and Higgs fields coupled to that group and to supergravity, here is a conjecture about the form and coefficient of this anomaly.  

\large 
\be
 \cA^{1}_{\rm Supergravity}=
\int d^4 x\; \ve^{\lam\m\n\s} \lt \{ c_1 C_{ \a}
\pa_{\lam}  \Y_{\m}^{\a}   \pa_{\n}  V^a_{\s} +\k c_2
\pa_{\lam}  \Y_{\m}^{\a}     \pa_{\n} \Y_{\s \a} \; \w^a \rt \} g^3 \k^3 M_G \lt < D^a\rt >
\la{conjecturedcodffofanomaly} 
\ee   

\Large 
Here we have written the coefficients $c_1,c_2$ so that they are dimensionless.
The reasoning behind  this conjecture resides simply in the above diagrams 
(\ref{anomdiagram}) and (\ref{offdiagonalprop}), as follows:

If we add group matrices to this paper rather than leave it in the Abelian notation, then we get the following group matrix in (\ref{anomdiagram}):
\be
\k^3 g^3M_G  M^2 v^i (T^{a} T^b T^a)_{i}^{\;\;j} {\ov v}_j
=
\lt (  \fr{1}{2} C_2(G)+ C_1(G) \rt )\k^3 g^3M_G  M^2 <D^a>
\ee
where $T^a$ are the representation matrices for a Lie group, and 
\be
<D^b>=M^2  v^i
 T_{i}^{b\;\;j} {\ov v}_j
;\; \lt <A^i \rt> = M v^i,
\ee
and we are taking into account what we get for the mixed propagator from 
(\ref{offdiagonalprop}).  We  note  that the vector propagator is diagonal in group space, if we drop the mass terms, and we can do that because we are interested only in the linearly divergent part of the diagram (\ref{anomdiagram}). But then we note that
\be
T^{a} T^b T^a
=
\lt ([T^{a}, T^b] + T^b T^a\rt ) T^a
=
 f^{abc} T^cT^a + T^b T^aT^a
\eb
=
 f^{abc} \fr{1}{2} f^{cad} T^d + T^b T^aT^a
= 
\lt (  \fr{1}{2} C_2(G)+ C_1(G) \rt )T^b
\ee
and so we get the form (\ref{conjecturedcodffofanomaly}), where we absorb this constant $\lt (  \fr{1}{2} C_2(G)+ C_1(G) \rt )$, made from Casimir coefficients of the group, into the coefficients $c_i$.  In \ci{freepro,Weinberg3} there are discussions about the question whether the auxiliary field $D^a$ can have a VEV, and \ci{Intriligator2007} is also useful.

 \refstepcounter{orange}
{\bf \theorange}.\;
\la{calculatingthediagrams}
{\bf Calculating the Diagrams:}  It would be nice to calculate the diagrams like (\ref{offdiagonalprop})  and (\ref{anomdiagram}).   The diagram (\ref{offdiagonalprop})  needs to be non-zero in order for the diagram  (\ref{anomdiagram}) to even exist.  But that requires us to have a proper version for both massive propagators, with gauge parameters,  in  (\ref{offdiagonalprop}).  That is why some effort has been made to start the discussion of the terms in (\ref{prop}) and
(\ref{Qprop}) for the gravitino propagator  in paragraph \ref{gravitinopropagatorterm}
 below.

 \refstepcounter{orange}
{\bf \theorange}.\;
\la{starttowritedowntheaction}
{\bf The Action for the model:}  
We take supergravity coupled to a U(1) super gauge theory, coupled to two   chiral multiplets ${\widehat A}_i = A_i + \q^{\a} \c_{i \a}+ \fr{1}{2}( \q^2) F_i, i = 1, 2$, with opposite charges, and two uncharged chiral multiplets  ${\widehat B}_i = B_i + \q^{\a} \f_{i \a}+ \fr{1}{2}( \q^2) G_i, i = 1, 2$ . We choose a simple superpotential of the 
O'Raiffertaigh \ci{west} form, so that the theory will have spontaneous breaking of supergravity.  This part can be done as if we were using merely rigid SUSY, and we use $M_O$ for a mass parameter here:

\be
\cA_{\rm Scalars\; from ;Chiral\; Superpotential }
\eb=\int d^4 x d^2 \q \lt \{
 g_1 {\widehat B}_1 
 {\widehat A}_1 {\widehat A}_{2 } 
 +{\widehat B}_2 \lt (  g_2 
  {\widehat A}_1 {\widehat A}_{2 }   
 -M_O^2 \rt )
\rt \}+ *
\eb
=\int d^4 x \lt \{
g_1  G_1 A_1 A_2
+ g_1 B_1 A_1   F_2   
+g_1
B_1  F_1    A_2   
\ebp
+ g_1  B_1 \c_{1}^{\a} 
\c_{2\a} 
+
g_1
\f_{1}^{\a}  \c_{1\a}   A_2 
+   g_1 \f_{1}^{\a}     A_1 \c_{2 \a} 
\ebp
+g_2  G_2 A_1 A_2
+ g_2 B_2 A_1  F_2    
+g_2 B_2  F_1 A_2   
\ebp
+ g_2  B_2 
\c_{1}^{\a} \c_{2\a} 
+   g_2 \f_{2}^{\a}  \c_{1\a}    A_2 
+   g_2  \f_{2}^{\a} A_1   \c_{2 \a}  
- G_2 M_O^2
\rt \}\ee
and we add to this the purely scalar terms relating to the auxiliary fields  and scalars in the other parts of the actions:

\large
\be
\cA_{\rm Scalars\; from\; Action}= \int d^4 s \lt \{\sum_{n=1}^2\lt \{ F_n \oF_n+
G_n \ovG_n\rt \} + D^2 +  D g   A_{1} \A_{1}  - D g   A_{2} \A_{2} 
\rt \}\ee
\Large
We put these together into 
\be
\cA_{\rm SP}=
\cA_{\rm Scalars\; from\; Action}
\eb
+\cA_{\rm Scalars\; from ;Chiral\; Superpotential }
+{\ov \cA}_{\rm Scalars\; from ;Chiral\; Superpotential }
\ee
\refstepcounter{orange}
{\bf \theorange}.\;
\la{eqsforausiliaries}
{\bf  The bosonic parts of the various auxiliary fields} here are obtained from the bosonic parts of their equations of motion:

\large
\be
 \fr{\d \cA_{\rm SP}}{\d G_1}  = \ovG_1 + g_1 
A_1 A_2=0
;\ \fr{\d \cA_{\rm SP}}{\d G_2} =  \ovG_2 +
 g_2 A_1 A_2 - M_O^2=0
\ee
 \be
  \fr{\d \cA_{\rm SP}}{\d F_1} = \ov F_1 +
  g_1   B_1 A_2  + g_2   B_2 A_2=0
;\;  \fr{\d \cA_{\rm SP}}{\d F_2} = \ov F_2 +
  g_1  B_1 A_1 + g_2  B_2 A_1=0
\ee
\be
 \fr{\d \cA_{\rm SP}}{\d D} =  D +
g
\lt (
A_{1}   \A_{1}
-A_{2}  \A_{2} 
\rt )
 =0
\ee

\Large
\refstepcounter{orange}
{\bf \theorange}.\;
\la{Vexpression}
{\bf  
We integrate the auxiliary fields in the path integral,} which replaces them with the negative of their values from their equations of motion, and this yields the following potential in the usual way: 
\be
V= -\cL_{\rm Scalar\; Potential }=\sum_{n=1}^2\lt \{ F_n \oF_n+
G_n \ovG_n\rt \}+ D D
\ee
which is, using the equations in paragraph \ref{eqsforausiliaries}: 
\be
V=  
 \lt | g_1   A_1 A_2 \rt |^2
+\lt |g_2  A_1 A_2  - M_O^2 \rt |^2
+
\lt | g_1   B_1 A_2  + g_2   B_2 A_2 \rt |^2
\eb
+
\lt |   g_1  B_1 A_1 + g_2  B_2 A_1 
\rt |^2
+   \lt | g 
\lt (A_{1}   \A_{1}
- \A_{2}A_{2}  \rt) 
\rt|^2
\la{potentialscalar}
\ee

\refstepcounter{orange}
{\bf \theorange}.\;
{\bf  
The minimum value of this expression occurs for Vacuum Expectation Values (VEVs) as follows: }
\be
\lt < B_1\rt >
=\lt < B_2\rt >=\lt < \oF_1\rt >
=\lt < \oF_2\rt >=0
\ee
These zero results all follow from the fact that setting them to zero just contributes zero to the VEV of the expression (\ref{potentialscalar}).  The VEVs of the auxiliary fields $G_1,G_2,D$ may be non-zero, which means that the
gravitino gains a mass, and that the supersymmetry here is spontaneously broken.  

\refstepcounter{orange}
{\bf \theorange}.\;
{\bf  Values of the VEVs:}  
\la{vevvaluesparagraph}
These VEVs need to satisfy the  equations obtained by taking the derivatives of the expression (\ref{potentialscalar}) with respect to the four scalar fields 
$A_{1} ,A_{2} ,B_{1} ,B_{2} $ and setting the result to zero, so as to find an extremum.
Then one chooses the minimum value for V among the extrema.  We can choose real values for the VEVs here.
The solution is  unique for present purposes, and is 
\be \lt < A_1\rt > 
=
\lt < A_2\rt >= M_O v = M_O
\sqrt{\fr{g_2}{g_1^2+ g_2^2}}
\eb \Ra
\lt < \ovG_1\rt >= - M_O^2 \fr{  g_1 g_2}{g_1^2+ g_2^2} ;\;
\lt < \ovG_2\rt >  =+ M_O^2 \fr{ g_1^2}{g_1^2+ g_2^2};\;
\lt < D \rt >= 0 
\la{valuesofvevs}
\ee
and it yields a minimum value for V, with those VEVs, of:
\be
V = V \lt[A_1 \ra \lt< A_1\rt >, A_2\ra\lt < A_2\rt > , B_1 \ra 0, B_2 \ra 0\rt ]=M_O^4 \fr{g_1^2  }{\lt (g_1^2 + g_2^2\rt)}
\la{valueofpotentialatminimum}
\ee

\refstepcounter{orange}
{\bf \theorange}.\;
{\bf  We expect to need a non-zero \cconst\ counterterm in any gravitational theory:} In supergravity theories this is of the anti-de-Sitter kind, negative in value, whereas the value (\ref{valueofpotentialatminimum}) is positive.  We can and do `fine tune' the above value to be equal to the \cconst, and so the result has zero \cconst\ corresponding to asssymptotically flat spacetime. That is the origin of the mass of the gravitino.

\refstepcounter{orange}
{\bf \theorange}.\;
\la{vectorbosonpropagator}
{\bf  
This is a bit different from the well known way that the mass of the vector boson appears in such theories:} In this case that mass occurs in the form
\be
-  D_{\m} A_1 \ovD^{\m} \A_1
- D_{\m} A_2 \ovD^{\m} \A_2 \ra
\eb
- i g M_O  v   \pa_{\m} V ^{\m}  \A_1+  i g M_O  v    \pa_{\m} V ^{\m}  A_1  
\eb
+ i g M_O  v   \pa_{\m} V ^{\m}  \A_2 
- i g M_O v   \pa_{\m} V ^{\m}  A_2 
\eb
- 2 g^2 M_O^2 v^2  V_{\m}  V^{\m}
\ee

\Large
\refstepcounter{orange}
{\bf \theorange}.\;
\la{cosmotogravitinomass}
{\bf  Cosmological term and  gravitino mass:}
In this model we need to suppose we start with a cosmological term.  
To preserve the BRS invariance of supergravity we also need to add a gravitino mass term and a term that changes the gravitino \tr.
\be
\cA_{\rm Cosmological}= \int d^4 x  \lt \{ e M_{\rm cc}^4  - e \lt (M_G \Y \Y - 
\fr{1}{e}M_{\rm New}^2 {\tilde \Y}^{\m} \s_{\m}\oC + *\rt ) \rt \}
\eb
=\int d^4 x \lt ( 1 +\fr{\k}{6} h^{\s}_{\;\;\s} +\cdots \rt )  \lt \{ M_{\rm cc}^4 
- \lt (M_G \Y \Y -\fr{1}{e} M_{\rm New}^2 {\tilde \Y}^{\m} \s_{\m}\oC + *\rt )
 \rt \}
\ee
and in order to have invariance we need
\be
M_G  M_{\rm New}^2 = \fr{\k}{6} M_{\rm cc}^4 
\ee

Now suppose that we have, using (\ref{valueofpotentialatminimum}) 
\be
M_O^4 \fr{g_1^2  }{\lt (g_1^2 + g_2^2 \rt)}=M_{\rm cc}^4 
\ee

so that the cosmological term $\int d^4 x \; M_{\rm cc}^4 e $ is cancelled by the VEV of V.
These terms have the right signs to accomplish this because the (uncancelled) cosmological term in supergravity corresponds to an anti de Sitter universe \ci{freepro}.
  We still have the gravitino mass term and we still have the new term in the transformation of the gravitino, but we no longer have the cosmological term in terms of $h$.  How does the supergravity symmetry survive?

Strictly speaking we really should integrate the auxiliary terms out at this point.  But for   present purposes we can proceed as follows.
We get several  new kinds of terms from the VEV of the auxiliary terms at the same time. We get the mixing terms from 
the lines that correspond to 
 (\ref{wherethegravitinomixingcomesfrom})
in $\cA_{\rm Chiral \;Kinetic\;B \;1}+\cA_{\rm Chiral \;Kinetic\;B \;2}$.  It is
\be
 \cA_{\rm Goldstino\;mixing}
=\int d^4 x  \lt \{
 \ov \f_1 \ov \s^{\m}
 \k <G_{1}>   
 + \ov \f_2 \ov \s^{\m}
 \k <G_{2}>    \rt \} \Y_{\m } 
\eb
=
\int d^4 x  M_{\rm New,3}\;
 \ov \f_{\rm Goldstino} \s^{\m}
    \Y_{\m } 
\la{goldstinomixing}\ee
Also we get the terms in the BRS transformations of the $\f$ fields in 
$\cA_{\rm ZJ \;Chiral ;B \;1}+ \cA_{\rm ZJ \;Chiral ;B \;2}$ from the lines that correspond to 
(\ref{termthatyieldsgoldstinoshiftterm}).  They are:
\be
\int d^4 x \lt \{
{\widetilde \f}_1^{\a} 
  \k <G_1>  C_{ \a} +{\widetilde \f}_2^{\a} 
  \k <G_2>  C_{ \a} +*
\rt\}
\eb
=
\int d^4 x \;M_{\rm New,4}\lt \{
{\widetilde \f}_{\rm Goldstino}^{\a} 
  C_{ \a}   
\rt\}+*
\ee

So now we have no cosmological term but we have the following special terms in the action

\be \int d^4 x \lt \{    M_G \Y \Y 
+ M_{\rm New,3}\;
 \ov \f_{\rm Goldstino} \ov \s^{\m}
    \Y_{\m } \rt \} + *
\ee
and special terms in the ZJ action too:
\be \int d^4 x \lt \{      - M_{\rm New}^2 {\widetilde \Y}^{\m} \s_{\m}\oC 
+M_{\rm New,4} 
{\widetilde \f}_{\rm Goldstino}^{\a} 
  C_{ \a}   
\ebp
+{\widetilde G}_1 
\oC_{\dot \b} {\ov \s}^{ \m \a \dot\b }  \k <G_1 >  \Y_{\m \a} 
+{\widetilde G}_2 
\oC_{\dot \b} {\ov \s}^{ \m \a \dot\b }  \k <G_2 >  \Y_{\m \a} 
 \rt \} + *
\ee

 There would also be a term $-i  \k 
{\ov\lam}^{\dot \b} {\ov \s}^{\m}_{ \dot \b \a}
  \lt < D \rt > \Y_{\m \a} + *
$ from line (\ref{originofgravitinomassfromgaugevev}) in the gauge action $\cA_{\rm Gauge}$  except that in  this model, from  paragraph \ref{vevvaluesparagraph}, we have $ \lt < D \rt >=0$
The Goldstino then is 
\be
{ \f}_{\rm Goldstino}=
\fr{\k}{2M_G} \lt ( \lt < \ovG_{1} \rt >   \f_1 
+   \lt < \ovG_{2} \rt >   \f_2 \rt )
=\fr{\k M_O^2}{M_G} \fr{  g_{2}}{g_1^2+ g_2^2}
\lt (  g_{2}  \f_1 
+ g_1    \f_2 \rt )
\ee

This needs more attention to see exactly how  the invariance remains, but we will not do that here  for now.

 \refstepcounter{orange}
{\bf \theorange}.\;
\la{gravitinomixpara}
{\bf  We can eliminate the Goldstino mixing term  $ \cA_{\rm Goldstino\;mixing}$ in (\ref{goldstinomixing})
by choosing a special form of the Ghost and Gauge Fixing Action:}
This technique was invented by 't Hooft for the case of the spontaneous breaking of the massive gauge vector boson, and later  it was updated using the BRS transformations \ci{taylor}. 
This also allows us to keep a  free (and useful)   gauge parameter.   We choose the following Ghost and Gauge Fixing (GGF) action for the gravitino:

\large
\be
\cA_{\rm Gravitino\; GGF} =
\int d^4 x 
 \d \lt \{
  E^{\a}   \lt (\Lam_{\a} + \lt ( \sqrt{m} \s^{\m}_{\a \dot \b}   {\ov \Y}_{\m}^{ \dot \b}
+  \sqrt{\fr{M_G^2}{m}} \f_{\a; \rm Goldstino} \rt )
\rt ) + *
\rt \}
\ee
\Large
where $E^{\a}$ is a Grassmann even antighost for the $C_{\a}$ SUSY ghost, and $\Lam^{\a}$ is a Grassmann odd `ghost auxiliary field', and these are taken to have the BRS variations: 
\be
\d E^{\a}=\Lam^{\a},\; 
\d  \Lam^{\a}=0
\ee
After the auxiliary $\Lam_{\a}$ and its complex conjugate are  integrated out of the path integral this becomes:

\large
 \be
\cA_{\rm Gravitino\; GGF} \Ra \int d^4 x \lt \{ - \lt ( \sqrt{m} \s^{\m}_{\a \dot \b}   {\ov \Y}_{\m}^{ \dot \b}
+  \sqrt{\fr{M_G^2}{m}} \c_{\a; \rm Goldstino} \rt ) 
\ebp
 \ve^{a \g} \lt ( \sqrt{m} \s^{\m}_{\g \dot \d}   {\ov \Y}_{\m}^{ \dot \d}
+  \sqrt{\fr{M_G^2}{m}} \c_{\g; \rm Goldstino} \rt ) 
\la{squareterm}
+ 
  E^{\a}  \d     \lt ( \sqrt{m} \s^{\m}_{\a \dot \b}   {\ov \Y}_{\m}^{ \dot \b}
+   \sqrt{\fr{M_G^2}{m}}\c_{\a; \rm Goldstino} \rt )  
\rt \}+ *
\ee
\Large
The cross term in line (\ref{squareterm}) cancels the mixing term (\ref{goldstinomixing}), 
  yielding a gauge parameter dependent mass term for the Goldstino, and a gauge fixing term for the Gravitino, along with a ghost action that involves the BRS variations of the Gravitino and the Goldstino. The gauge parameter is $m$ and the Gravitino mass is $ M_{\rm G} $. Both have the dimension of mass.

All of this is exactly like what happens for the massive vector boson and its Goldstone boson and its gauge parameter, except that the gauge parameter of the Gravitino has to have the dimension of mass. The old style of gauge fixing the gravitino  \ci{kallosh} did not accomodate any gauge parameter.

\refstepcounter{orange}
{\bf \theorange}.\;
{\bf The Gravitino Propagator with a gauge parameter}
\la{gravitinopropagatorterm}
Calculating the form of this with a gauge parameter is not simple.  There are ten `spinor-tensor-derivative' operators that must be considered. 
Here is the kinetic action which results after the mixing term (\ref{goldstinomixing})
has been removed using the methods in paragraph \ref{gravitinomixpara}:
 
\be
\cA^{0}_{\rm Kinetic\;Gravitino}= \int d^4  x\;
\Y_{\m} \g^0
K^{\m \n}\Y_{\n}  
\eb
=  
\int d^4  x\;
\Y_{\m} \g^0
 \lt (
   \ve^{\m\n\r\k}\g_{\r} \g_5  \pa_{\k}
 - M_G
   \h^{\m\n} 
 - m
   \g^{\m} 
       \g^{\n} 
 \rt )
 \Y_{\n} 
       \ee
 Here we have switched  to real gamma matrices as used in \ci{wessandzumino}. 
This term is a  Lorentz covariant  tensor in the Lorentz indices $\m \n \cdots $, and a  Lorentz covariant  matrix in the implicit spinor indices that go with $\g_{\m} \cdots$. and we want to find the propagator, which is the inverse:
\be
P^{\m \n}K_{\n \lam}
=
 \d^{\m}_{\;\;\lam}
 \ee
 
 To find this inverse we need to write down the general Lorentz covariant form of this tensor-matrix.  This turns out to be:
\large\[
P^{\m \n}=
\lt \{ 
\lt (
a_1\ve^{\m\n}_{ \;\;\;\a \b} \g^{\a}\pa^{\b} \g_5  
+a_2 \pa^{\m}  \g^{\n} 
+a_3  \pas \h^{\m \n} 
+a_4   \g^{\m}  \pa^{\n} 
+
 \fr{c_1}{m^2} \pa^{\m} \pas \pa^{\n} 
\rt ) 
\rt.
\]
\be
\lt.
+
\lt (
 \fr{b_1}{m} \pa^{\m}  \pa^{\n} 
+ \fr{b_2}{m} \Box \h^{\m \n}
+ \fr{b_3}{m} \Box \g^{\m} \g^{\n}
+ \fr{b_4}{m} \g^{\m} \pas \pa^{\n}
+ \fr{b_5}{m} \pa^{\m} \pas \g^{\n}
\rt )
 \rt \}
\la{genform}
\ee

\Large
In (\ref{genform}) we need to assume that the ten coefficients $a_1,\cdots c_1,b_1 \cdots$ are functions of the ratio
\be
R= \fr{\Box}{m^2}
\ee
which has mass dimension zero.

So far, this work has only been done for the case where $M_G=0$. For that case, 
the solution is:
\be
P^{\m \n}=\fr{1}{2 \Box} 
\lt \{ 
Q^{\m\n}
+
 \pa^{\m}  \pa^{\n}
 \fr{2 }{m }
 \rt \}
\la{prop}
\ee
where
\be
Q^{\m\n}
=\ve^{\m\n}_{ \;\;\;\a \b} \g^{\a}\pa^{\b} \g_5  
+  \pa^{\m}  \g^{\n} 
-   \pas \h^{\m \n} 
+   \g^{\m}  \pa^{\n} 
- \pa^{\m}  \pa^{\n}
  \fr{4 \pas }{ \Box}   
\la{Qprop}
\ee
Note that the propagator has a simple form:
\bitem
\item
When $m\ra 0$ this becomes singular. There is no inverse when $m=0$.
\item
There are no $\fr{1}{\Box + m^2}$ poles.   The only place that $m$ appears in the propagator is in the term
$\fr{1}{m } \fr{\pa^{\m}  \pa^{\n}}{\Box}$ which looks and acts like a gauge parameter term.
 
\item
The part
$ Q^{\m \n}$ satisfies $\g_{\m}  Q^{\m \n}=0$
\item
The part $   \ve^{\m\n\r\k}\g_{\r} \g_5  \pa_{\k}$ satisfies 
\be
   \ve^{\m\n\r\k}\g_{\r} \g_5  \pa_{\k}
   \pa_{\n}
=0
\ee

\end{itemize}

For the case where $M_G\neq 0$ this still needs to be completed.  
 Obviously, this propagator needs to be used so that one can calculate diagrams like 
(\ref{offdiagonalprop})--the significant physics of that diagram needs to be gauge invariant under two different gauge parameters. The physical consequences, including the anomaly coefficient, of the diagram (\ref{anomdiagram}) also should be gauge invariant under the gauge parameter for the massive vector boson.

\refstepcounter{orange}
{\bf \theorange}.\;
{\bf  The Action and the BRS transformations for the model used in this paper:}  
At this point we need to write down the  complicated and long action and transformations for the theory used in this paper.  These are needed to understand the remarks in \ref{newcounterterms} to see what is going on. The invariance of the action and the nilpotence of the BRS \tr s are all summarized in the Master Equation, found in the next section.

\refstepcounter{orange}
{\bf \theorange}.\;
\la{mastereq}
{\bf The Master Equation:} This summarizes all the consequences of the BRS symmetry of the theory.  For the total action $\cA$, which is the sum of all the actions and all the ZJ actions, it takes the form:

\large

\be
\cM_{\rm Total}=
\cM_{\rm Sugra}
+
\cM_{\rm Gauge}
+
\cM_{\rm A\;Chiral; \;1}
\eb
+
\cM_{\rm A\;Chiral; \;2}
+
\cM_{\rm B\;Chiral; \;1}
+
\cM_{\rm B\;Chiral; \;2}
=0
\la{masterequation}\ee
where
\be
\cM_{\rm Sugra}
= \int d^4 x \lt \{
\fr{\d \cA}{\d h^{\m a} }
\fr{\d \cA}{\d {\widetilde h}_{\m a} }
+\fr{\d \cA}{\d \Y^{\m \a} }
\fr{\d \cA}{\d {\widetilde \Y}_{\m \a} }
+\fr{\d \cA}{\d { {\ov\Y}}^{\m \dot\a}  }
\fr{\d \cA}{\d {\widetilde {\ov\Y}}_{\m \dot\a} }
\ebp
+\fr{\d \cA}{\d{w}_{\m a b}}
\fr{\d \cA}{\d{\widetilde w}_{\m a b}  }
+\fr{\d \cA}{\d \r^{ab}}
\fr{\d \cA}{\d {\widetilde \r}_{ab} }
+\fr{\d \cA}{\d \x^{\m}}
\fr{\d \cA}{\d {\widetilde \x}_{\m} }
+\fr{\d \cA}{\d C^{\a}}
\fr{\d \cA}{\d {\widetilde C}_{\a} }
+\fr{\d \cA}{\d \oC^{\dot \a}}
\fr{\d \cA}{\d {\widetilde \oC}_{\dot \a} }
\rt \} 
\ee

\be
\cM_{\rm Gauge}
= \int d^4 x \lt \{
\fr{\d \cA}{\d V^{\m} }
\fr{\d \cA}{\d {\widetilde V}_{\m} }
+\fr{\d \cA}{\d \lam^{\a} }
\fr{\d \cA}{\d {\widetilde \lam}_{\a} }
+\fr{\d \cA}{\d \ov \lam^{\dot \a} }
\fr{\d \cA}{\d {\widetilde {\ov \lam}}_{\dot \a} }
+\fr{\d \cA}{\d D}
\fr{\d \cA}{\d {\widetilde D} }
+\fr{\d \cA}{\d \w}
\fr{\d \cA}{\d {\widetilde \w} }
\rt \} 
\ee

\be
\cM_{\rm A\;Chiral; \;n=1,2}
= \int d^4 x \lt \{
\fr{\d \cA}{\d A_n }
\fr{\d \cA}{\d {\widetilde A}_{n} }
+\fr{\d \cA}{\d \A_n }
\fr{\d \cA}{\d {\widetilde {\A}}_{n} }
\ebp
+\fr{\d \cA}{\d \c_{n}^{ \a} }
\fr{\d \cA}{\d {\widetilde \c}_{n \a} }
+\fr{\d \cA}{\d { {\ov \c}_n}^{ \dot\a}  }
\fr{\d \cA}{\d {\widetilde {\ov\c}}_{n \dot\a} }
+\fr{\d \cA}{\d{F}_{n }}
\fr{\d \cA}{\d{\widetilde F}_{n}  }
+\fr{\d \cA}{\d {\ov F}_n}
\fr{\d \cA}{\d {\widetilde {\ov F}}_{n} }
\rt \} 
\ee

\be
\cM_{\rm B\;Chiral; \;n=1,2}
= \int d^4 x \lt \{
\fr{\d \cA}{\d B_n }
\fr{\d \cA}{\d {\widetilde B}_{n} }
+\fr{\d \cA}{\d \oB_n }
\fr{\d \cA}{\d {\widetilde {\oB}}_{n} }
\ebp
+\fr{\d \cA}{\d \f_{n}^{ \a} }
\fr{\d \cA}{\d {\widetilde \f}_{n \a} }
+\fr{\d \cA}{\d { {\ov \f}_n}^{ \dot\a}  }
\fr{\d \cA}{\d {\widetilde {\ov\f}}_{n \dot\a} }
+\fr{\d \cA}{\d{G}_{n }}
\fr{\d \cA}{\d{\widetilde G}_{n}  }
+\fr{\d \cA}{\d {\ov G}_n}
\fr{\d \cA}{\d {\widetilde {\ov G}}_{n} }
\rt \} 
\ee
\Large
The Master Equation actually applies to the 1PI vertex functional $\cG = \cA + \hbar \cG_1 + O(\hbar^2)$, and the action $\cA$ is just the local part of that. The next sections will contain the various actions which sum to the total action  $\cA$ .  There are also some simple adjustments needed for complications that relate to the gauge fixing terms and ghost actions.  We  will ignore those complications for now.

The BRS ZJ $\d$ operator is the `square root' of the above, by which we mean
\be
\d = \int d^4 x \lt \{
\fr{\d \cA}{\d h^{\m a} }
\fr{\d  }{\d {\widetilde h}_{\m a} } 
+
\fr{\d \cA}{\d {\widetilde h}_{\m a} } 
\fr{\d  }{\d h^{\m a} }
+ \cdots\rt \}
\ee
Note that the nilpotence of  $\d$ and the vanishing of  $\cM_{\rm Total}
$  imply each other: 
\be
\cM_{\rm Total}=0 \Lra \d^2=0
\ee
It should also be remembered that $\cM$ is rather like a Poisson Bracket, and so it is invariant under transformations that are like canonical transformations \ci{goldstein,LandauLifmechanics}.  That is what is happening with the generation of the new counterterms in paragraph \ref{generator}, as was explained in \ci{dixonnucphys}.

\refstepcounter{orange}
{\bf \theorange}.\;
{\bf  The Action and the BRS transformations for the Four Chiral Kinetic Multiplets, Coupled to Gauge Theory, and to  Supergravity:}  
\la{chiralA1action}
Here is the action for the theory used in this paper:

\be
\cA_{\rm Chiral \;Kinetic\;A \;1}\equiv \cA_{A_1}=
\eb 
\int d^4 x e \lt \{
\lt (   \pa_{\m}  A_{1}   -i g  V_{\m} A_{1}     
 +\k \Y_{\m} \cdot  \c_{1} \rt ) 
\lt (   \pa^{\m}  \A_{1}    + i g   V^{\m} \A_{1}    
 +\k {\ov \Y}^{\m} \cdot  \ov\c_{1} \rt ) 
\la{wherethebosonmasscomesfrom}
\ebp
+ \ov \c_1 \ov \s^{\m}
\lt [ \pa_{\m}  \c_{1 \a}    - i g     V_{\m}  \c_{1 \a}    + w_{\m ab} (\s^{ab})_{\a \b} \c_{1}^{ \b}
+  \k F_{1}   \Y_{\m \a}  \rt.
\la{wherethegravitinomixingcomesfrom}
\ebp
\lt.
 +\k \s^{\t}_{\a \dot \b}  \ov \Y_{\m}^{\dot \b}  
\lt (   \pa_{\t}  A_{1}    - i g  V_{\t} A_{1}     
 +\k \Y_{\t} \cdot  \c_{1} \rt )
\rt ]
 + * - F_{1} \oF_{1 } 
\ebp
+  g \lt ( 
D A_{1} \A_{1} +\lam^{\a}  
\c_{1 \a} \A_{1 }
+ \ov \lam^{\dot \a} 
{\ov \c}_{1 \dot \a} A_{1}
\rt )
\rt \}
\la{lastlineofchiralaction}
\ee

The action for  $\cA_{A_2}$ is obtained, from 1 above, by simply changing $1 \ra 2$ and also changing the sign of g throughout. The actions for $\cA_{B_1}$ and $\cA_{B_2}$ are the same except for the change of notation (the superfields are discussed above, and indicate the notation) and setting $g\ra 0$, since they are neutral under the gauge group. In the above we are using  `supercovariant derivatives', as discussed in \ci{west}.  We have written these out in full  in the first line (\ref{wherethebosonmasscomesfrom}) as
 $ {\widehat D}_{\m} A_{1}  {\widehat D}^{\m} \A_{1} $, for example.

\refstepcounter{orange}
{\bf \theorange}.\;
\la{sugraaction}
{\bf  The Action for  
Supergravity:}  Here is the  Action for supergravity as used here:
\be
{\cL}_{\rm Supergravity}
=
 M_P^2 e R 
+  \ve_{\m\n \lam \s}  \Y^{\m \a} \s^{\n}_{\a \dot \b} \pa^{\lam} \ovY^{\s \dot \b}
+  \ve_{\m\n \lam \s}  \Y^{\m } \s^{\n} w^{\lam cd} \s_{cd} \ovY^{\s }
\ee
We use the formulation by Zumino and Deser in \ci{zuminodeser}. There is an important feature here that relates to the spectral sequence and to the quantization of these theories which we discuss in paragraph \ref{firstandsecondpara}.

\refstepcounter{orange}
{\bf \theorange}.\;
\la{U1action}
{\bf  The Action for the U(1) Gauge Theory, Coupled  to 
Supergravity:}  

\be
\cA_{\rm Gauge}= 
\int d^4 x \lt \{
- \fr{1}{4} 
\lt (
F_{\m \n} 
+\k \Y_{[\m}^{\g}\s_{\n] \g \dot \d} \ov \lam^{  \dot \d}
+\k\ov\Y_{[\m}^{\dot \g} \ov \s_{\n] \dot \g  \d} \lam^{\d}  
\rt )
\ebp
\lt (F^{\m \n} 
+ \k\Y^{[\m}_{\ve}\s^{\n] \ve \dot \z} \ov \lam_{ \dot \z}
+\k\ov\Y^{[\m}_{\dot \e} \ov \s^{\n] \dot \e  \z} \lam_{\z}  
\rt ) \rt \}
\ee
\be
+\int d^4 x  \lt \{
{\ov\lam}^{\dot \b} {\ov \s}^{\m}_{ \dot \b \a}
\Bl [
 \pa_{\m} \lam^{\a}+
 w_{\m}^{ab}
( { \s}_{a b})^{\a \b} 
 \lam_{\b} 
+
\k( \s_{\s\t})^{\a\b} F^{ \s \t} \Y_{\m\b}  
- i D \k \Y_{\m \a}
\Br ]
\la{originofgravitinomassfromgaugevev}
\ebp
+({\rm*\; of\; previous \;term})
 - D^2
\rt \}
\ee
where
we define
\be
F_{\m \n} =
\pa_{\m} V_{\n} - \pa_{\n} V_{\m}
\ee

\refstepcounter{orange}
{\bf \theorange}.\;
{\bf  The ZJ part of the Supergravity Action:} 
\la{ZJaction}
This is also needed for the theory.  It describes the nilpotent BRS transformations that leave the action invariant.
\large
\be
{\cL}_{\rm ZJ,  Supergravity} = 
 {\widetilde h}^{\m a} 
\lt \{
- M_P( \pa_{\m} \x_{a}  )
+
M_P(   \r_{\m a})
+
 \r_{a}^{\;\; b} h_{\m b}
 -h_{\m \s}  \pa^{\s} \x_{a}   
\ebp
+\x^{\s} \pa_{\s}  h_{\m a} 
+ \oC \s_{a} \Y_{\m}
+  C \ov \s_{a} \ov \Y_{\m}
+ \k \lt (
\Y_{\m}
 \s^{\n} \oC  
+
C \s^{\n}\ov \Y_{\m }
\rt )
h_{\n a}
\rt \}
+ {\widetilde w}^{\m a b} 
\lt \{
\pa_{\m} \r_{ab} 
\ebp
+ [ \r, w_{\m} ]_{ab}
+\x^{\s} \pa_{\s}  w_{\m a b}
+ \k B_{\m ab}
+ \k \oB_{\m ab}
+ \k^2 \lt ( B^{c}_{\;\;\; ac}h_{\m b}- B^{c}_{\;\;\; bc}h_{\m a} \rt )
\ebp
+ \k^2 \lt ( \oB^{c}_{\;\;\; ac}h_{\m b}- \oB^{c}_{\;\;\; bc}h_{\m a} \rt )
- \k \lt (
\Y_{\m}
 \s^{\n} \oC  
+
C \s^{\n}\ov \Y_{\m }
\rt )
w_{\n a b}
\rt \}
\ee
\be
+ {\widetilde \Y}^{\m } 
 \lt \{
M_P( \pa_{\m} C  
+
w_{\m  ab}\s^{ab}   C )+  \r^{ab} \s_{ab}\Y_{\m}
\ebp
+
 h_{\m\n} \pa^{\n} C  
+\x^{\s} \pa_{\s}  \Y_{\m }
- \k \lt (
\Y_{\m}
 \s^{\n} \oC  
+
C \s^{\n}\ov \Y_{\m }
\rt )
\Y_{\n }
\rt \}
\ee
\be
+ {\widetilde \x}^{a } 
\lt \{
- C^{\a}  \s_{a \a \dot \b} \oC^{\dot \b}
+ \k C \s^{\m} \oC h_{ \m a}
+ \x^{\s} \pa_{\s} \x_{a} 
 \rt \}
\ee
\be
 + {\widetilde \r}^{ab }  \lt \{
(\r \r)_{ab}
+ \x^{\s} \pa_{\s} \r_{ab} 
+ \k C \s^{\m} \oC w_{\m ab }
\rt \}
\ee
\be
+ {\widetilde C}_{\a} 
\lt \{
\r^{ab}
(\s_{ab})^{\a\b}  C_{\b} + \x^{\s} \pa_{\s} C^{\a}
+ \k C \s^{\m} \oC \Y_{\m}^{\;\; \a}
 \rt \}
\ee
\be
+ {\widetilde \oC}_{\dot \b}
\lt \{
\r^{ab}\ov \s_{ab}  \oC  
+ \x^{\s} \pa_{\s} \oC
+ \k C \s^{\m} \oC \ov \Y_{\m}
\rt \}^{\dot \b}\ee
\Large
where following Zumino and Deser in \ci{zuminodeser} we put
\be
B_{a}^{\;\;\lam \m}=
 \ve^{\lam \m \n \r} \oC \s_{a} D_{\n} \y_{\r}
\ee

\refstepcounter{orange}
{\bf \theorange}.\;
{\bf  The ZJ part of the Gauge Action:} 
\la{Gaugeaction}
This  describes the nilpotent BRS transformations of the gauge ZJ sources, that leave the action invariant.

\be
\cA_{\rm ZJ, Gauge}= 
\int d^4 x \lt \{
{\widetilde V}^{\m} \lt ( \pa_{\m} \w  + C^{\a} \s_{\m \a \dot \b} {\ov \lam}^{\dot \b}
\rt.\ebp\lt.
 + \oC^{\dot \a} \ov \s_{\m \dot \a  \b}  { \lam}^{\b} 
+\x^{\s} \pa_{\s} V_{\m}
\rt)
\ebp
+{\widetilde \lam}^{\a} \lt (  \s_{\a \b}^{\m\n} {\widehat F}_{\m \n}  C^{\b} 
+ \r_{ab} \s^{ab}_{\a \b}  \lam^{\b}
+
 i D C_{\a} 
+\x^{\m} \pa_{\m} \lam^A_{\a}
 \rt )
\ebp
+{\widetilde {\ov \lam}}^{\dot \a} \lt (
{\ov\s}_{\dot\a \dot\b}^{\m\n} {\widehat F}_{\m \n}  \oC^{\dot\b} 
 + \r_{ab} \ov\s^{ab}_{\dot\a \dot\b}  \lam^{\dot\b} 
- i D \oC_{\dot \a}+\x^{\m} \pa_{\m} {\ov \lam}_{\dot \a } 
\rt)
\ebp
+{\widetilde D}\lt ( i( C^{\a} \s_{\a \dot \b}^{\m} {\widehat {\ov D}}_{\m} {\ov \lam }^{\dot \b}  
 - \oC^{\dot \a} \ov \s_{\dot \a  \b}^{\m}  {\widehat D}_{\m} { \lam }^{\b} )
 +\x^{\m} \pa_{\m} D\rt)\ebp
+{\widetilde \w}\lt (C^{\a} \s_{\a \dot \b}^{\m} {\oC}^{\dot \b} 
V_{\m} +\x^{\m} \pa_{\m}  \w\rt)
\rt \}
\ee
In the above we are using `supercovariant derivatives', indicated with the notation 
$ {\widehat D}_{\m} { \lam }^{\b} $, for example,  as discussed in \ci{west} and as used above in paragraph \ref{U1action}.

\refstepcounter{orange}
{\bf \theorange}.\;
{\bf  The ZJ part of the Four Chiral Actions:} 
\la{GaugeactionZJ}
This  describes the nilpotent BRS transformations of the four kinds of chiral  ZJ sources, that leave the action invariant.
\be
\cL_{\rm ZJ \;Chiral\;A \;1}=\eb 
{\widetilde A}_1 \lt [ -ig \w  A_1   +C \cdot \c_1 +  \x^{\m} \pa_{\m } A_1
\rt ]
\eb
+{\widetilde \A}_1 \lt [+ ig \w  A_1   +\oC \cdot {\ov\c}_1 +  \x^{\m} \pa_{\m } \A_1
\rt ]
\eb
+{\widetilde \c}_1^{\a} \lt [
 -i g \w  \c_{1 \a}+ \r^{ab} (\s_{ab})_{\a \b}   \c_{1}^{ \b}
+   F_1  C_{ \a} 
+ \x^{\m} \pa_{\m }  \c_{1\a} 
\rt.
\ee
\be
\lt.+ \s^{\t}_{\a \dot \b}  \oC^{\dot \b}  
 \lt ( \pa_{\t} A_1 - i g V_{\t} A_1      +\k    \Y_{\t}^{\b}  \c_{1 \b} \rt )
\rt ]+* 
\la{termthatyieldsgoldstinoshiftterm}
\ee
\be
+{\widetilde F}_1 \lt [
  -ig  \w  F_{1} +\x^{\m}   \pa_{\m } F_1     
\rt ]+* 
\eb
+{\widetilde F}_1 
\oC_{\dot \b} {\ov \s}^{ \m \a \dot\b }
\lt [   \pa_{\m}  \c_{1 \a}   
-ig  V_{\m}  \c_{1_\a}   
+  \k F_1   \Y_{\m \a} 
\rt ]+* 
\eb
+{\widetilde F}_1 
\oC_{\dot \b} { \s}^{ \m \a \dot\b }
\k \s^{\t}_{\a \dot \g}  \ov \Y_{\m}^{\dot \g}  
 \lt [ \pa_{\t} A_1 - i g V_{\t} A_1   
   +\k    \Y_{\t}^{\b}  \c_{1 \b} \rt ]
+* 
\ee

The expressions for $\cL_{\rm ZJ \;Chiral\;A \;2}$ and $\cL_{\rm ZJ \;Chiral\;B \;n},n=1,2$ are obtained from the above in the same way that was described above for the actions.

\refstepcounter{orange}
{\bf \theorange}.\;
\la{generator}
{\bf  The Diagram (\ref{offdiagonalprop}) can exist only if the gravitino is massive:} 
All these counterterms will be expected to be obtained from an expression like this (the meaning of the notation is given below).  Note that the ghost charge of this expression is 
$N_{\rm Ghost} = -1$: it is linear in ZJ sources:
\be
\cA^{-1}_{\rm Generator}= \int d^ 4 x  \lt \{ i f_1 {\widetilde D} \lt ( F_1- \oF_1\rt )
+i f_2 {\widetilde D} \lt ( F_2- \oF_2\rt )
\ebp
 + f_3  \lt ( 
{\widetilde \c}_{1}^{\a} \lam_{\a} 
+ {\widetilde {\ov \c}}_{1}^{\dot \a} {\ov \lam}_{\dot \a} 
\rt ) + f_4  \lt ( 
{\widetilde \c}_{2}^{\a} \lam_{\a} 
+ {\widetilde {\ov \c}}_{2}^{\dot \a} {\ov \lam}_{\dot \a} 
\rt )\ebp
 + i f_5 \lt ( {\widetilde F}_1 - {\widetilde \oF}_1 \rt ) D
 + i f_6 \lt ( {\widetilde F}_2 - {\widetilde \oF}_2 \rt ) D
+\ebp
 + f_7  \lt (
{\widetilde \lam}^{\a}  \c_{1\a} 
+ {\widetilde {\ov \lam}}^{\dot \a} {\ov \c}_{1\dot \a} 
\rt ) + f_8  \lt (
{\widetilde \lam}^{\a}  \c_{2\a} 
+ {\widetilde {\ov \lam}}^{\dot \a} {\ov \c}_{2\dot \a} 
\rt ) +\cdots \rt \}
\ee
To get the counterterms one takes
\be
\d \cA^{-1}_{\rm Generator}=\cA^{0}_{\rm Counterterms}
\ee
where the operator $\d$ comes from the `square root'
 of the Master equation, found in equation  (\ref{masterequation}).  
It is easy to verify that there are plenty of diagrams that appear when the gravitino is massive, and we expect them all to be related in this way.  It is interesting to note that these counterterms change the scalar interaction in paragraph \ref{starttowritedowntheaction} in a simple way.  That does not appear to spoil the general argument or the VEVs here, however.

\refstepcounter{orange}
{\bf \theorange}.\;
{\bf Errors:} Anyone who works with SUSY makes plenty of these, and the author is no exception.  There are so many minus signs to take care of that it is exhausting, and short cuts must be taken.  If the reader thinks there is a missing minus sign or factor or term, he may well be right.  Some features are crucial of course, but they are usually simple.  The crucial and simple issues in this paper are in paragraphs 
 \ref{sugranomalyform}
and \ref{sugracounter1}, and in the diagrams (\ref{anomdiagram}) and
(\ref{offdiagonalprop}). The author  apologizes for not yet having the energy to carefully and accurately check all the terms in the actions and transformations.  I believe that these are pretty close to correct, but there are hundreds of terms to check, and the task is not easy.  There are two stages in finding the action and transformations here.  The first stage is to make sure that there are enough terms of the right kinds, with indices in the right places, to satisfy  each identity.  The second stage is to make sure that the signs and factors are correct.  I believe that I have accomplished the first stage, but I have currently not yet had the energy to perform the second stage. In the case of $\d w_{\m ab}$, I am still working on the first stage.  Much of the necessary material there is already in the paper \ci{zuminodeser}.  However the current results reported here are, I think, sufficient to justify the conjecture  that spontaneous breaking is needed to generate these sugranomalies.    What difference does this make to this paper?  I  think that the fundamental result is not affected at all.  It  involves only the simple terms discussed above.  The calculation of the anomaly coefficient and the question whether it requires spontaneous breaking of supersymmetry are crucially important, of course.  To calculate the anomaly coefficient we need the exact action and BRS transformations.  We need to be sure that the Master equation is correct and that $\d^2=0$ is correct, and that no mistakes are made in the calculation. That is a large undertaking.

\refstepcounter{orange}
{\bf \theorange}.\;
{\bf First Order and Second Order Formalism:} \la{firstandsecondpara}
Most work in supergravity has been done using the second order formalism, where the spin connection $w_{\m a b}$ is replaced by its expression in terms of the vierbein $e_{\m a} = \h_{\m a} + \k h_{\m a}$.  We do not do that here for two reasons.  The first is that it seems easier to formulate the BRS formalism in terms of the first order formalism.  The second is that it seems very strange to try to quantize the theory when $w_{\m}^{a b}$  is no longer present.  It  is the gauge field of the ghost $\r^{ab}$. This transformation is the usual sort of gauge transformation:   $\d w_{\m}^{a b}=\pa_{\m} \r^{a b}+ \cdots $.  The absence of $w_{\m}^{a b}$  means that in some sense, in the second order formalism, one needs to have two different antighosts for the gauge fixing term for the field  $  h_{\m a}$, one for the general coordinate ghost $\x_{\m}$ and the other for the Lorentz ghost $\r^{ab}$.  From the spectral sequence point of view, this is particularly disturbing and it looks wrong. We also prefer the Weyl two component complex formalism because the Fierz transformations are so much simpler there.  It seems clear that supersymmetry is more natural, and simpler,  in that formalism. For example the anomaly (\ref{sugranomaly}) discussed here is very chiral--its complex conjugate 
\be
{\ov \cA}^{1}_{\rm Supergravity}= \int d^4 x\; \ve^{\lam\m\n\s} \lt \{ \oc_1 \oC_{\dot \a}
\pa_{\lam}  {\ov \Y}_{\m}^{\dot \a}   \pa_{\n}  V_{\s} + \oc_2
\pa_{\lam}  {\ov \Y}_{\m}^{\dot \a}     \pa_{\n} {\ov \Y}_{\s \dot \a} \; \w + \cdots\rt \}
\ee 
is also an anomaly form.

\refstepcounter{orange}
{\bf \theorange}.\;
{\bf Auxiliary Fields:}
This paper is composed using the first order formalism
and in terms of Weyl two component complex spinors. The reader might wonder whether we need additional auxiliary fields to close the algebra of supergravity, for example in the expressions in paragraph \ref{ZJaction}.  Auxiliary fields are useful for generating a calculus where multiplets can be multiplied to make other multiplets and invariants.  However they are not needed to create the nilpotent BRS transformations.  The reason is that one can imagine having the auxiliary fields and integrating them out using the path integral \ci{Dixon:1991wt}.  
That changes the terms in the Master Equation, but the derived $\d$ is still nilpotent, because it must be nilpotent, given that it is the `square root' of the Master Equation.  As is well known, auxiliary fields are used to close transformations that do not close, except using the field equations.  In the Master Equation approach, exactly the same function is served by the Zinn sources. It appears to me that there is no need in the first order formalism for any auxiliary fields in the present paper--they would show up as a need for terms with doubled Zinn sources. But it seems likely that they would start to appear at higher orders, in particular  because of the terms in paragraph \ref{generator}.

\refstepcounter{orange}
{\bf \theorange}.\;
{\bf Spectral Sequence:}
The results in this paper were   obtained by the author using spectral sequences 
\ci{Dixon:1990jv,Dixon:1991wi,Dixon:1991my,Dixon:1993jtmin,Dixon:1993yj}.
  These methods provided the insight needed to deduce that there had to be an anomaly in supergravity.  When a suitable spectral sequence can be used, the space $E_{\infty}$ represents all the possible invariants of the theory--but their explicit construction is still needed in that case. In the present situation, a spectral sequence indicated that there ought to be an anomaly, but its construction from that took a very long time. 
Unfortunately, those methods are quite difficult and confusing, though powerful.  There are still lots of unsolved issues there when those methods are applied to supergravity, and indeed other theories too.

\refstepcounter{orange}
{\bf \theorange}.\;
{\bf Future Steps:} 
The action and the transformations, and their closure, need to be completed and rechecked. The work here on the massless gravitino propagator with a massive gauge parameter needs to be extended to the case where the gravitino is also massive.
The supergravity anomalies need to be calculated, in many theories.
The values of their coefficients should be independent of all gauge parameters. The significance  of these supergravity anomalies needs to be understood.  Is the conjecture about their coefficients in equation (\ref{conjecturedcodffofanomaly})  correct?  If it is correct, does that mean that the viable theories must be those which do have some $\lt <F^i \rt >\neq 0$, but which also have the sugranomaly coefficient, or  perhaps $\lt <D^a \rt >$ itself, arranged to vanish, somehow? Is that easy?  What kinds of theories are those? The spectral sequence work also needs to be advanced.

\vspace{.3cm}

\refstepcounter{orange}
{\bf \theorange}.\;
{\bf Supergravitational fields in this model:} 
\normalsize
\begin{center}
\begin{tabular}{| c c c c c |}
\hline     Symbols & 
\begin{tabular}{ c }
Ghost\\ Number\\
\end{tabular}  &\begin{tabular}{ c }
Reality and\\ Grassman\\
\end{tabular}  & \begin{tabular}{ c }
 Dim $m^{d}$\\ $d=$\\
\end{tabular}   & Name    \\ \hline
$  w_{\m}^{\;\;ab}$ &0 &  Real, Even& 1     & Lorentz  Gauge Field\\  
$\r^{ab}$  &1  & Real, Odd &      $   0 $                 & Lorentz Tensor Ghost\\
$\z^{ab}$  &-1  & Real, Odd &      $   2 $                 & Lorentz Tensor AntiGhost\\
$R^{ab}$  &0  & Real, Odd &      $   2 $                 & Lorentz  Ghost Auxiliary\\
$ \Y_{\m}^{\;\;\a},  \ovY_{\m}^{\;\;\dot \a}  $   &0 & Complex, Odd  & $ \fr{3}{2}$                           & Gravitino Gauge Field\\
$C^{\a}, \oC^{\dot \a}$ &1   &  Complex, Even &      $   - \fr{1}{2} $                 &Gravitino  Spinor Ghost\\
$E^{\a}, \oE^{\dot \a}$ &-1   &  Complex, Even &      $   \fr{5}{2} $                 &Gravitino  Spinor AntiGhost\\
$\Theta^{\a}, {\ov \Theta}^{\dot \a}$  &0  &  Complex, Even &      $   \fr{5}{2} $                 &Gravitino Ghost Auxiliary\\
$e_{\m a} \ra \h_{\m a} + \k h_{\m a} $ &0   & Real, Even &      $      0 $                 & Vierbein\\
$h_{\m a}$ &0   & Real, Even & 1                           & Gravitational Field\\
$\x_{\m}$ &1   & Real, Odd &      $   - 1 $                 & General Coordinate  Ghost\\
$\z_{\m}$  &-1  & Real, Odd &      $   3 $                 & General Coordinate  AntiGhost\\
$H_{\m}$ &0   & Real, Odd &      $   3 $                 & General Coordinate  Ghost Auxiliary\\
\hline
\end{tabular}
\end{center}

\Large
Note that we define:
\be
e_{\m a} = \h_{\m a} + \k h_{\m a} ;
g_{\m \n} = e_{\m a} \h^{ab} e_{\n b}\ee
which means that the inverse of $e$ is an infinite series in $h$, and I am ignoring that for now.

\refstepcounter{orange}
{\bf \theorange}.\;
{\bf Gauge Particles for the Abelian Gauge Theory:} 
\normalsize

\begin{center}

\begin{tabular}{| c c c c c |}
\hline     Symbols &\begin{tabular}{ c }
Ghost\\ Number\\
\end{tabular} &\begin{tabular}{ c }
Reality and\\ Grassman\\
\end{tabular}  & \begin{tabular}{ c }
 Dim $m^{d}$\\ $d=$\\
\end{tabular}   & Name  \\
\hline
  \\ \hline
$V_{\m}$&0  & Real, Even & 1                           & Gauge Vector Boson\\
$\w $ &1  & Real, Odd & 0                           & Gauge Ghost Field\\
$\h $ &-1  & Real, Odd &         2                   & Gauge AntiGhost Field\\
$H $&0   & Real, Odd & 2                           & Gauge Ghost Auxiliary Field\\
$\lam^{ \a}, \ov \lam^{A 
\dot \a}$&0   & Complex, Odd & $\fr{3}{2}$                           & Gauge Fermion Field\\
$D$&0   & Real, Even & 2                           & Gauge Auxiliary Field\\
\hline
\end{tabular}
\end{center}

\Large
\refstepcounter{orange}
{\bf \theorange}.\;
{\bf 
Higgs Representations in this Model}
\normalsize

\begin{center}

\begin{tabular}{| c c c c |}
\hline     Symbols &\begin{tabular}{ c }
Reality and\\ Grassman\\
\end{tabular}  & \begin{tabular}{ c }
 Dim $m^{d}$\\ $d=$\\
\end{tabular}   & Name  \\
\hline
  \\ \hline
$A_n,\A_{n} $  & Complex, Even & 1                    &  Charged Chiral Scalar Pair: n=1,2\\
$\c_n^{\a}, \ov \c_{n }^{\dot \a}$  & Complex, Odd & $\fr{3}{2}$  & 
Charged Chiral Spinor Pair :n=1,2\\
$F_n,\oF_{n}$  & Complex, Even & 2                           &  Charged Chiral Auxiliary Pair:n=1,2\\
\hline
$B_n ,\B_{n} $  & Complex, Even & 1                           &  Singlet Chiral Scalars: n=1,2
\\$\f_n^{\a}, \ov \f_{n }^{\dot \a}$  & Complex, Odd & $\fr{3}{2}$    &  Singlet Chiral Spinors :n=1,2\\
$G_n,\ovG_{n}$  & Complex, Even & 2                           &  Singlet Auxiliary Scalars: n=1,2\\
\hline
\end{tabular}
\end{center}

\Large
\refstepcounter{orange}
{\bf \theorange}.\;
{\bf 
Some Constants in this Model}

\normalsize
\begin{center}
\begin{tabular}{| c c c c |}
\hline     Symbols &\begin{tabular}{ c }
Paragraph 
\end{tabular}  & \begin{tabular}{ c }
 Dim $m^{d}$\\ $d=$\\
\end{tabular}   & Name    \\ \hline
$\k= \fr{1}{M_{{}_{P}}} $  & &      $   - 1 $                 & Planck Length\\
$M_{{}_{P}} = \fr{1}{\k} $  &   &      $   +1 $                 & Planck Mass\\
$M_{O}   $  & \ref{starttowritedowntheaction} &      $   +1 $                 & 0'Raiffeartaigh Mass\\
$M_{G}   $  &\ref{cosmotogravitinomass},\ref{gravitinopropagatorterm}  &      $   +1 $                 & Gravitino Mass\\
$m   $  & \ref{gravitinopropagatorterm}  &      $   +1 $                 & Gravitino Gauge Parameter\\
$M_{{}_{\rm CC}}^4  $  &\ref{cosmotogravitinomass}
&      $   +4 $                 &Cosmo Mass\\
$g_{1}   $  & \ref{starttowritedowntheaction} &      $   0 $                 &  Chiral Coupling\\
$g_{2}  $  & \ref{starttowritedowntheaction} &      $   0 $                 & Chiral Coupling \\
$g    $  & \ref{starttowritedowntheaction} &      $   0 $                 & Gauge Coupling\\
$v =\sqrt{\fr{g_2}{g_1^2+ g_2^2}}$  &  \ref{vevvaluesparagraph} &      $   0 $                 & VEV parameter\\
$M_{\rm 1,mix}= g <A_1>=g v M_O   $  & \ref{chiralA1action},\ref{diagramforanomalyparagraph}
  &      $   1 $                 & Mass term from VEV\\
$\lt < A_1\rt > = M_O v $  & \ref{vevvaluesparagraph} &      $   1 $                 & VEV of $A_1$\\
$\lt < A_2\rt > = M_O v $  &  \ref{vevvaluesparagraph} &      $   1 $                 & VEV of $A_2$\\
$\lt < G_1\rt > = - M_O^2 \fr{  g_1 g_2}{g_1^2+ g_2^2}  $  &  \ref{vevvaluesparagraph}  &      $   2 $                 & VEV of $G_1$\\
$\lt < G_2 \rt > =M_O^2 \fr{ g_1^2}{g_1^2+ g_2^2};
$  &  \ref{vevvaluesparagraph}  &      $   2 $                 & VEV of $G_2$\\
\hline
\end{tabular}
\end{center}
\Large

\refstepcounter{orange}
{\bf \theorange}.\;
{\bf 
Some Important Concepts in the paper}
\normalsize
\begin{center}
\begin{tabular}{| c c c  |}
\hline     Symbol &\begin{tabular}{ c }
Paragraph 
\end{tabular}  
   & Name  and Comment  \\ 
\hline$\cM$ &\ref{mastereq} & Master Equation\\
$\d$ &\ref{mastereq} & BRS Nilpotent Transformation\\
$\widetilde{\rm F}$ &\ref{mastereq} & ZJ Source for the BRS Variation $ \d {\rm F} $ of the Field F: \\
$\cA^{1}_{\rm Supergravity}$ & \ref{sugranomalyform} & Supergravity Anomaly\\
$\cA^{0}_{\rm Supergravity\;Counterterm\;2}$ &\ref{sugracounter1}
& Generates boundary for Supergravity Anomaly\\
$V= -\cL_{\rm Scalar\; Potential }$&\ref{Vexpression}&Scalar\; Potential \\
${ \f}_{\rm Goldstino}$ &\ref{cosmotogravitinomass}
& Goldstino Fermion\\
$\cA^{-1}_{\rm Generator}$&\ref{generator}
 &Generator for new terms in action\\
$ P_{\m\n}, Q_{\m\n}$& 
\ref{gravitinopropagatorterm}
&(\ref{prop}) , (\ref{Qprop})  Gravitino Propagator\\

\hline
\end{tabular}
\end{center}

\begin{center}
 {\bf Acknowledgments}
\end{center}
\vspace{.1cm}

\Large

  I thank Doug Baxter, Carlo Becchi, Philip Candelas,  James Dodd, Mike Duff, Richard Golding,  Dylan Harries,   Pierre Ramond, Graham Ross, Peter Scharbach,    Kelly Stelle,  Xerxes Tata,  J.C. Taylor, and Peter West for stimulating correspondence and conversations.

\vspace{1cm}\tiny\numberhere \hspace{.2cm}\articlenumber\\\today
\end{document}